\documentclass[12pt]{article}
\usepackage{epsf}
\title{ Quark distributions in  QCD sum rules:  unexpected features and paradoxes} \vspace{2cm}
\author{ A.G.Oganesian\\
\\
Institute of Theoretical and Experimental Physics\\
B.Cheremushkinskaya 25, 117218, Moscow, Russia}
\date{}
\begin{document}

\maketitle

\newcommand{\be}{\begin{equation}}
\newcommand{\ee}{\end{equation}}

\def\la{\mathrel{\mathpalette\fun <}}
\def\ga{\mathrel{\mathpalette\fun >}}
\def\fun#1#2{\lower3.6pt\vbox{\baselineskip0pt\lineskip.9pt
\ialign{$\mathsurround=0pt#1\hfil##\hfil$\crcr#2\crcr\sim\crcr}}}

\begin{abstract}
\noindent  Some very unusual features of the hadron structure
functions, obtained in the generalized QCD sum rules, like the
surprisingly strong difference between longitudinally and
transversally polarized $\rho$ mesons structure functions and  the
strong suppression of the gluon sea in longitudinally polarized
$\rho$ mesons are discussed. Also the problem of exact zero
contribution of gluon condensates to pion and longitudinally
polarized $\rho$ meson quark distributions is discussed.
\end{abstract}

It is well known, that in QCD  DGLAP equations predicts the
evolution of these distributions with $Q^2$ but not the initial
values at some relatively low $Q^2$ from which this evolution
starts . To obtain these initial values (inputs) usually any form
of quark (valence and sea) and gluon distributions with some free
parameters are assumed. Then by using DGLAP equations, quark and
gluon distributions are calculated at all $Q^2$ and $x$ and
compared with the whole set of the experimental data, and in this
way the parameters are fixed. Evidently, such an approach is not
completely satisfactory from theoretical point of view, because,
we can loose some unexpected physical features of distributions,
so  it would be desirable to determine the initial distribution
directly from QCD. This talk will be devoted to the problem of the
determination of quark and gluon distributions in hadrons in model
independent way in QCD, and we will see, that some unexpected
features of quark distributions really appear.

A method to determinate valence quark distribution in QCD sum
rules at intermediate $x$ was suggested in \cite{1} and
generalized in \cite{2}-\cite{3}. Due to this generalization it
became possible to calculate valence quark distributions in pion
\cite{2}  and transversally and longitudinally polarized
$\rho$-meson \cite{3}, and also significantly improve the result
for valence quark distributions in proton \cite{4}.

Let me at first briefly remind the method. We start from
consideration of the 4-point correlator corresponding to the
forward scattering of hadron and electromagnetic currents.

\be
 \Pi = -i \int~ d^4x d^4y d^4z e^{ip_1x + iqy - ip_2z} \langle 0
\vert T \left \{j^h(x),~ j^{el}(y),~ j^{el}(0),~ j^h(z) \right \}
\vert 0 \rangle \ee

Here $p_1$ and $p_2$ are the initial and final momenta carried by
hadronic the current $j^h$, $q$ and $q^{\prime} = q + p_1 - p_2$
are the initial and final momenta carried by virtual photons
(Lorentz indices are omitted) and $t = (p_1 - p_2)^2 = 0$.

The idea of the approach (in the improved version) is to consider
the imaginary part (in $s$-channel) of this four-point correlator
$\Pi(p_1, p_2, q, q^{\prime})$. It is supposed that virtualities
of the photon $q^2, q^{\prime 2}$ and hadron currents $p^2_1,
p^2_2$ are large and negative $\vert q^2 \vert = \vert q^{\prime
2} \vert \gg \vert p^2_1 \vert,~ \vert p^2_2 \vert \gg R^{-2}_c$,
where $R_c$ is the confinement radius. It was shown in \cite{1}
that in this case the imaginary part in $s$-channel $[s = (p_1 +
q)^2]$ of $\Pi(p_1, p_2; q_1, q^{\prime})$ is dominated by a small
distance contribution at not small $x$, so OPE is applicable.

To find structure function one should compare dispersion
representation of the forward scattering amplitude in terms of
physical states with those in OPE and use Borel transformation.
Though we should consider the case of forward scattering, i.e.
$p_1=p_2$, but, and it is significant point of the method, we
should keep $p_1^2$ not equal to $p_2^2$ in all intermediate
stages and only in final result take the forward scattering limit.
Only in such way  it is possible to effectively (exponentially)
suppress all terms in sum rules expect those which are
proportional to structure function and achieve good accuracy.
Finally equating the physical and QCD representations (for detail
see \cite{2}) after double borelization (on $p_1^2$ and $p_2^2$)
we found

\be Im~ \Pi^0_{QCD} + \mbox{Power~ correction} = 2 \pi F_2 (x,
Q^2) g^2_h e ^{-m^2_h(\frac{1}{M^2_1} + \frac{1}{M^2_2})} \ee

Here $Im \Pi^0_{QCD}$ is the bare loop contribution, (where
continuum contribution is eliminated) and $g_h$ is defined as
$\langle 0 \vert j_h \vert h \rangle = g_h $. In what follows, we
will suppose $M^2_1 = M^2_2 \equiv 2 M^2$.

Let us start with the case of the pion. To find the pion structure
function one should choose the imaginary part of 4-point
correlator  with two axial (hadron) and two electromagnetic
currents and consider the invariant amplitude at tensor structure,
$P_{\mu} P_{\nu} P_{\lambda} P_{\sigma}/\nu$, where
$P=(p_1+p_2)/2$, ${\mu},{\nu}$ are vector current indexes and
${\lambda}, {\sigma}$ -hadron current indexes. I shall briefly
note the main points of calculations, for detail see \cite{2}. In
QCD part of sum rules we take into account the following terms:

1. Unit operator contribution (bare loop, Fig.1)  and leading
order (LO) perturbative corrections proportional to
$ln(Q^2/\mu^2)$ , where $\mu^2$ is the normalization point. In
what follows, the normalization point will be chosen to be equal
to the Borel parameter $\mu^2 = M^2_1=M^2_2=2M^2$.

2.  Power corrections -- higher order terms of OPE. Among them,
first of all we should account the dimension-4 correction,
proportional to the gluon condensate $\langle 0 \vert
\frac{\alpha_s}{\pi} G^n_{\mu \nu}~ G^n_{\mu\nu} \vert 0 \rangle$.
Corresponding diagrams are shown on Fig.2.  But, surprisingly,  it
was found that these diagrams contribution to the sum rule is
exactly cancelled after Borel transformation on $p_1^2$ and
$p_2^2$. I want to note, that diagrams in total may not cancel,
but only that part of them, which have imaginary part both $p_1^2$
and $p_2^2$ (i.e. just those which contribute to the pion
structure function).  So we see there is no gluon condensate
contribution to the sum rule for pion structure function. The fact
of exact cancellation of contributions of 10 diagrams on fig.2  is
surprising, but of course one can threat this as a just accidence.
But just the same take place for operators of the next dimension!

There are a large number of loop diagrams, corresponding to $d=6$
corrections. First of all, there are diagrams which correspond to
interaction only with gluon vacuum field, i.e. only with external
soft gluon lines (see Fig.3). They are, obviously, proportional to
$\langle g^3 f^{abc} G^a_{\mu \nu} G^b_{\alpha \beta} G^c_{\rho
\sigma} \rangle$. But one should also take into account  diagrams
of Fig.4 which are proportional to $\langle 0 \mid D_{\rho}
G^a_{\mu \nu} D_{\tau}G^a_{\alpha \beta} \mid 0 \rangle$ and
$\langle 0 \mid  G^a_{\mu \nu} D_{\rho} D_{\tau }G^a_{\alpha
\beta} \mid 0 \rangle$ . Using the equation of motion it can be
expressed in the terms of

$$
[\langle 0\mid g \bar{\psi}\psi\mid 0\rangle ]^2 ~~ \mbox{and} ~~
\langle 0 \mid g^3 G^a_{\mu \nu} G^b_{\nu\rho} G^c_{\rho \mu}
f^{abc} \mid 0 \rangle
$$

As it was shown in \cite{2}, if one take the sum of all diagrams,
then the terms, proportional to gluon condensate $\langle 0 \mid
g^3 G^a_{\mu \nu} G^b_{\nu\rho} G^c_{\rho \mu} f^{abc} \mid 0
\rangle$ exactly cancel each other just in the same way, as for
$d=4$ operator we have discussed above. I want to note, that there
is about a hundred different diagrams, so it is hard to believe
that such a cancellation is again an accidence.

So we can fixed first very unexpected feature of pion dynamics  -
gluon vacuum contribution to pion valence quark distribution is
exact zero for dimensions 4,6. I want to add, that, as we will see
a little later, just the same effect appear for longitudinally
polarized $\rho$ meson, so this is not the specific feature of
pion. It is very like that here we meet a consequence of some
symmetry, though today we cannot understand it.

Of course except the diagrams on fig. 2a,b there are another
diagrams of dimension 6, directly proportional proportional to the
four-quark operators $a^2=\alpha_s (2\pi)^4 (\langle 0 \vert
\bar{\psi} \psi \vert 0 \rangle)^2 $. I will not discuss them in
this talk, one can see papers \cite {2}, \cite {3}. Finally quark
distribution function (hereafter we will note it as QDF) has the
following form:

$$
xu_{\pi}(x) = \frac{3}{2\pi^2}\frac{M^2}{f^2_{\pi}}x^2(1-x) \Biggl
[ \Biggl ( 1+ \Biggl (\frac{a_s(\mu^2)\cdot
ln(Q^2_0/\mu^2)}{3\pi}\Biggr )$$


\be \times N(x)(1-e^{-s_0/M^2})-\frac{4\pi \alpha_s(\mu^2)\cdot
4\pi a^2}{(2\pi)^4 \cdot 3^7\cdot 2^6\cdot M^6} \cdot
\frac{\omega(x)}{x^3(1-x)^3}\Biggr ], \label{19} \ee where
$N(x)=(1-x)(1+4xln(1-x))-2x(1-2x)lnx~$ and $\omega(x)$ is the
fourth degree polynomial in $x$,
 This function
$u_{\pi}(x)$ may be used as an initial condition at $Q^2 = Q^2_0$
for solution of QCD evolution equations. The result is shown on
fig.5 (thick line).

In the numerical calculations we choose $Q^2_0 = 2 ~GeV^2$ (for
pion) and $a^2=0.23GeV^6$, found recently from $\tau$ decays
analysis -see  \cite{iz}, \cite{giz}, \cite{ii}. The continuum
threshold was varied in the interval $0.8 < s_0 < 1.2GeV^2$ and it
was found, that the results depend only slightly on it's
variation. The analysis of the sum rule shows, that it is
fulfilled in the region $0.2 < x < 0.7$, where both power
correction and continuum contribution are less than $30\%$ and the
stability in the Borel mass parameter $M^2$ dependence in the
region $0.4 GeV^2 < M^2 < 0.7 GeV^2$ is good. The accuracy of this
result was estimated in \cite {2} to be $20-30\%$. Comparison with
various experimental result (see \cite {2}) lead to conclusion,
that agreement with experiment is good. I want to note, that this
result for QDF is based only on the QCD sum rules.

It is possible also to estimate the second moments of QDF. Assume,
that $u_{\pi}(x) \sim 1/\sqrt{x}$ at small $x \la 0.15$ according
to the Regge behavior and $u_{\pi}(x) \sim (1-x)^2$ at large $x
\ga 0.7$ according to quark counting rules. Then, matching these
functions with our result, one may find the numerical values of
the first and the second moments of the $u$-quark distribution:
${\cal{M}}_1 = \int \limits^1 _0 u_{\pi} (x) dx \approx 0.84$,
${\cal{M}}_2 = \int \limits^1_0 xu_{\pi} (x) dx \approx 0.21$
where the values various slightly at reasonable changes of QDF
behavior at large or small $x$. The moment ${\cal{M}}_1$ has the
meaning of the number of $u$ quarks in $\pi^+$ and it should be
${\cal{M}}_1 = 1$. The deviation of $M_1$ from 1 characterizes the
accuracy of our calculation. The moment ${\cal{M}}_2$ has the
meaning of the pion momentum fraction carried by the valence $u$
quark. Therefore, the valence $u$ and $\bar{d}$ quarks carry about
40\% of the total momentum what is close to experimental results.

In the same way one can calculate valence $u$-quark distribution
in the $\rho^+$ meson. The choice of hadronic current is evident.
$j_{\mu}^{\rho} = \overline{u}\gamma_{\mu}d$ and the matrix
element is $\langle \rho^+\mid j^{\rho}_{\mu}\mid 0 \rangle =
\frac{m^2_{\rho}}{g_{\rho}}e_{\mu}$
 where $m_{\rho}$ is the $\rho$-meson mass, $g_{\rho}$ is the
$\rho-\gamma$ coupling constant, $g^2_{\rho}/4\pi=1.27$ and
$e_{\mu}$  is the $\rho$ meson polarization vector.

In the non-forward amplitude the tensor structure for
determination of $u$-quark distribution in the longitudinal $\rho$
meson is to $P_{\mu} P_{\nu} P_{\sigma}P_{\lambda}$, while for
transverse $\rho$ it is $-P_{\mu}P_{\nu}\delta_{\lambda\sigma}$)
(see \cite{3}).

In the case of longitudinal $\rho$ meson the tensor structure,
that is separated  is the same as in the case of the pion. It was
shown in \cite {3},  that $u$-quark distribution in the
longitudinal $\rho$ meson can be found from Eq.(3) by substituting
$m_{\pi}\to m_{\rho}$, $f_{\pi}\to m_{\rho}/g_{\rho}$. Sum rules
for $u^L_{\rho}(x)$  are satisfied in the wide $x$ region:  $0.1 <
x < 0.85$ with high accuracy (about $10\%$). Fig.5 (curve with
squares) presents $xu^L_{\rho}(x)$ as a function of $x$. The
values $M^2=1$ GeV$^2$ and $s_0=1.5$ GeV$^2$, $Q_0^2=4$ GeV$^2$
were chosen. It is clear, that in this case again gluon condensate
contribution is cancelled. If we construct the moment of QDF for
$\rho_L$ in the same way as for pion then we found $M_1=1.06$ and
$M_2=0.39$ (for one valence quark). One should note, that accuracy
of this prediction of moment of $\rho_L$ is very high, because in
this case sum rules for QDF are covered almost all $x$ region, as
I have noted before, so extrapolation procedure contribution
numerically is negligible.

So we can fix the second very unusual effect: the valence quarks
carry about $80\%$ of the total momentum  of longitudinally
polarized $\rho$ meson, so gluon sea should be strongly suppressed
(less than 20$\%$).

Let us now consider the case of transverse $\rho$-meson, i.e., the
term proportional to the structure
$P_{\mu}P_{\nu}\delta_{\lambda\sigma}$. The procedure of
calculations are the same except one significant points.

1. In contrast to the pion case, the $\langle
G^a_{\mu\nu}G^a_{\mu\nu}\rangle$ correction for transversally
polarized $\rho(\rho_T)$, .

2. In contrast to $\pi$ and $\rho_L$-meson cases, the terms
proportional to  $\langle G^a_{\mu\nu}G^a_{\mu\nu}\rangle$, as
long as  $\langle 0\mid  g^3 f^{abc} G^a_{\mu\nu}G^b_{\nu\rho}$
$G^c_{\rho\mu} \mid 0 \rangle$  are not cancelled for $\rho_T$.
But $\langle 0 \mid g^3 G^a_{\mu\nu} G^b_{\nu\rho}
G^c_{\rho\mu}f^{abc} \mid 0 \rangle$ is not well known; so we need
to use estimations based there are only some estimates based on
the instanton model. Result for QDF in transversally polarized
$\rho$ meson is shown on Fig.5 for the region $0.2<x<0.65$, (line
with asters). One can find complete analytical form and it detail
analysis in our paper \cite{3}. The choose of parameters are the
same as in previous cases. The main sources of inaccuracy are
$d=6$ gluon condensate (gives about $20\%$) and $d=4$ gluon
condensate (due the inaccuracy of it own value about factor 1.5,
which lead to uncertainty in QDF about $20\%$). Accuracy is worse,
about $30-50\%$ (better in the middle of region and worse at the
end). From fig.5 one can see, that difference between QDF in the
longitudinally ($\rho_L$) and  transversally ($\rho_T$) polarized
$\rho$ mesons is very large, many times larger that uncertainties.

So, and this is third surprise, we can conclude that QDF
significantly depend of polarization. Such great difference
between QDF of $\rho_L$ and $\rho_T$ is very difficult to explain
in usual quark model. I want to add, that  to check this effect,
we calculate the second moment  of QDF of $\rho_L$ and $\rho_T$ in
quite another way, considering of 2-point correlator  in external
tensor field. This method was offered in \cite {bb} for pion case
many years ago (for the method itself of the QCD sum rules in the
external fields see \cite {is}). In the \cite {sam} we use this
method for polarized $\rho$ meson case. The results totally
confirm the conclusion that QDF of $\rho_L$ and $\rho_T$ are quite
differ : the second moment $M_2$ for one $\rho_L$ is about 84$\%$
(compare with  78$\%$ discussed above) while $M_2$ for $\rho_T$ is
much smaller and close to usual hadron case (about 40$\%$ -
50$\%$). (More detail this results and comparison of this two
methods one can found in \cite {sam}.) That why I think, that QDF
indeed strongly depend on polarization, at least in $\rho$-meson
case, and this fact can't be explained in the present quark models
of the hadrons.

Summarizing, in this talk I tried to pay your attention to very
strange results, which appear from pure QCD non-model calculation
of quark distribution function and yet have no satisfactory
physical explanation. These are:

a) First of all, the total cancellation of diagrams with gluon
condensate (for dimensions $d=4,6$) seems to note to some unknown
symmetry. Of course it is interesting to check this property also
for dimension $d=8$, but it is technically very hard and isn't
complete yet.

b) Second, such a strong difference between QDF for longitudinally
polarized  ($\rho_L$) and transversally polarized ($\rho_T$) , as
is shown on fig.5, can not be explained in the usual hadron quark
models.

Taking in mind also the third very strange fact that, as we
discuss before, for $\rho_L$ gluon sea is strongly suppressed
(less than 15$\%$), we can conclude quark dynamic in ($\rho_L$)
and ($\rho_T$) should be quite different, in contradiction to
usual quark models. From my point of view it is extremely
interesting to understand this strange features, at least in some
model way.

Author is thankful to B.L.Ioffe for useful discussions.

This work was supported in part by CRDF grant RUP2-2621-MO-04 and
RFFI grant 06-02-16905.

\newpage
\begin{figure}
\epsfxsize=5cm \epsfbox{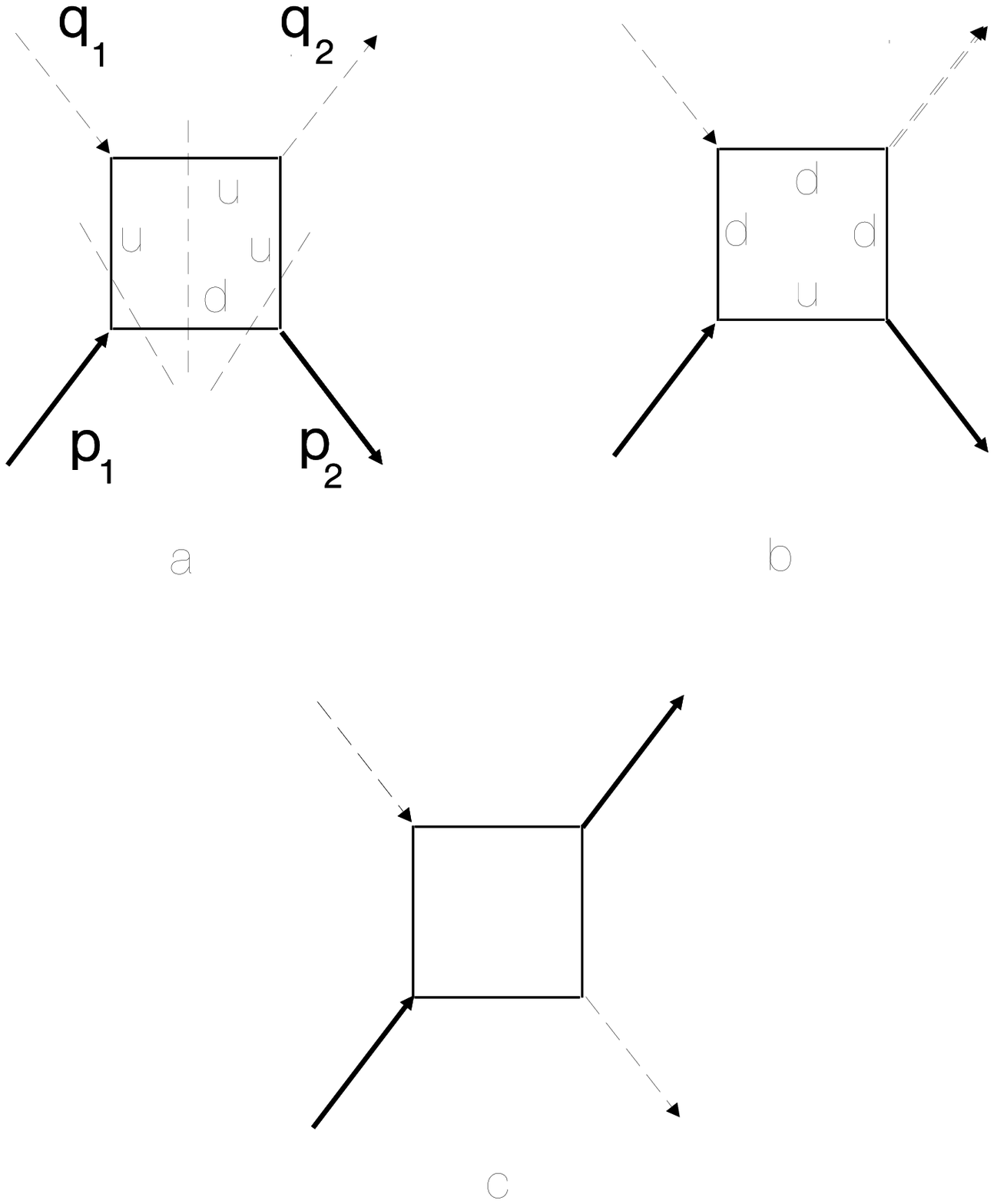} \caption{diagrams, corresponding
to bare loop contribution}
\end{figure}

\begin{figure}
\epsfxsize=5cm \epsfbox{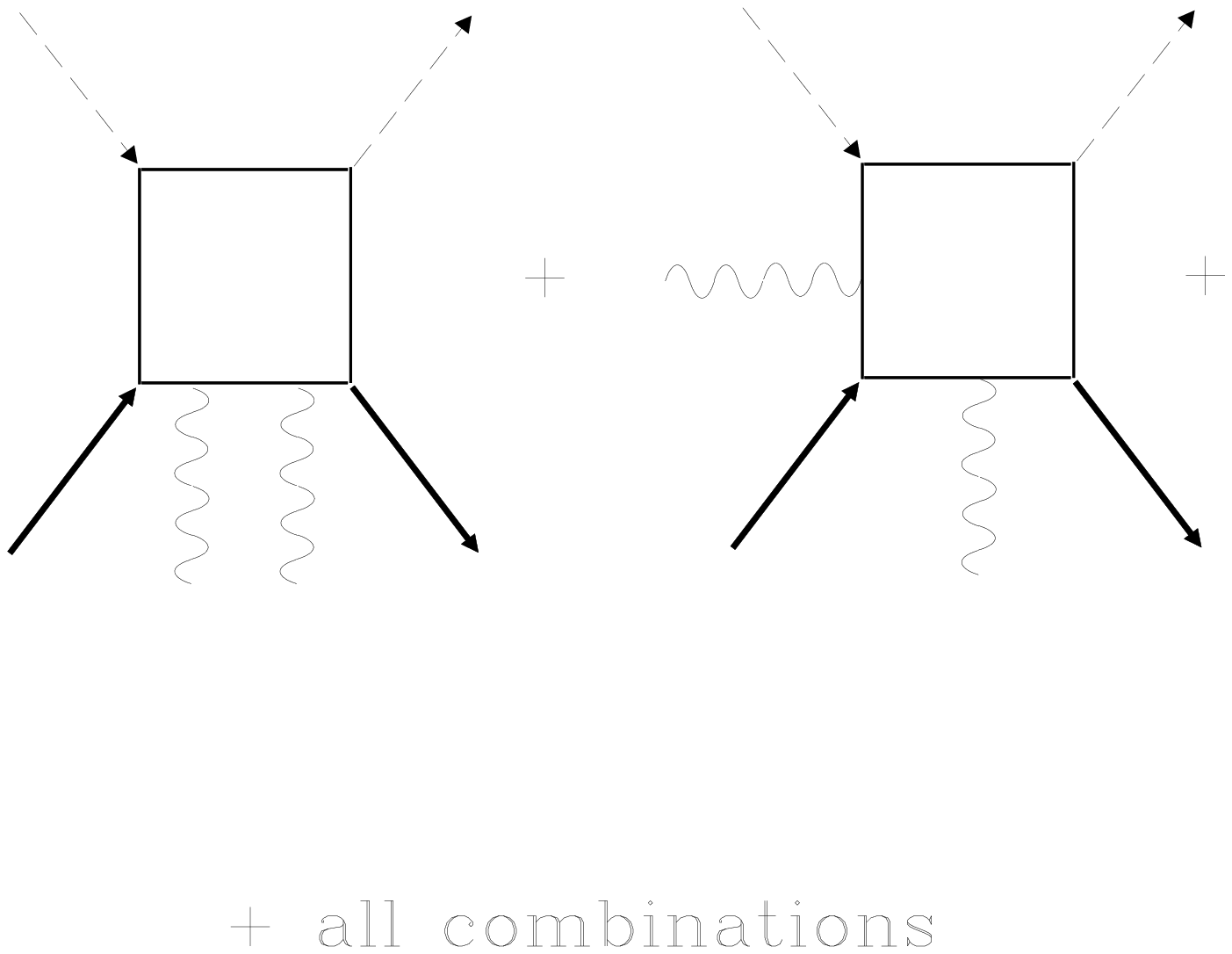} \caption{diagrams, corresponding
to $d=4$ gluon condensate}
\end{figure}

\begin{figure}
\epsfxsize=5cm \epsfbox{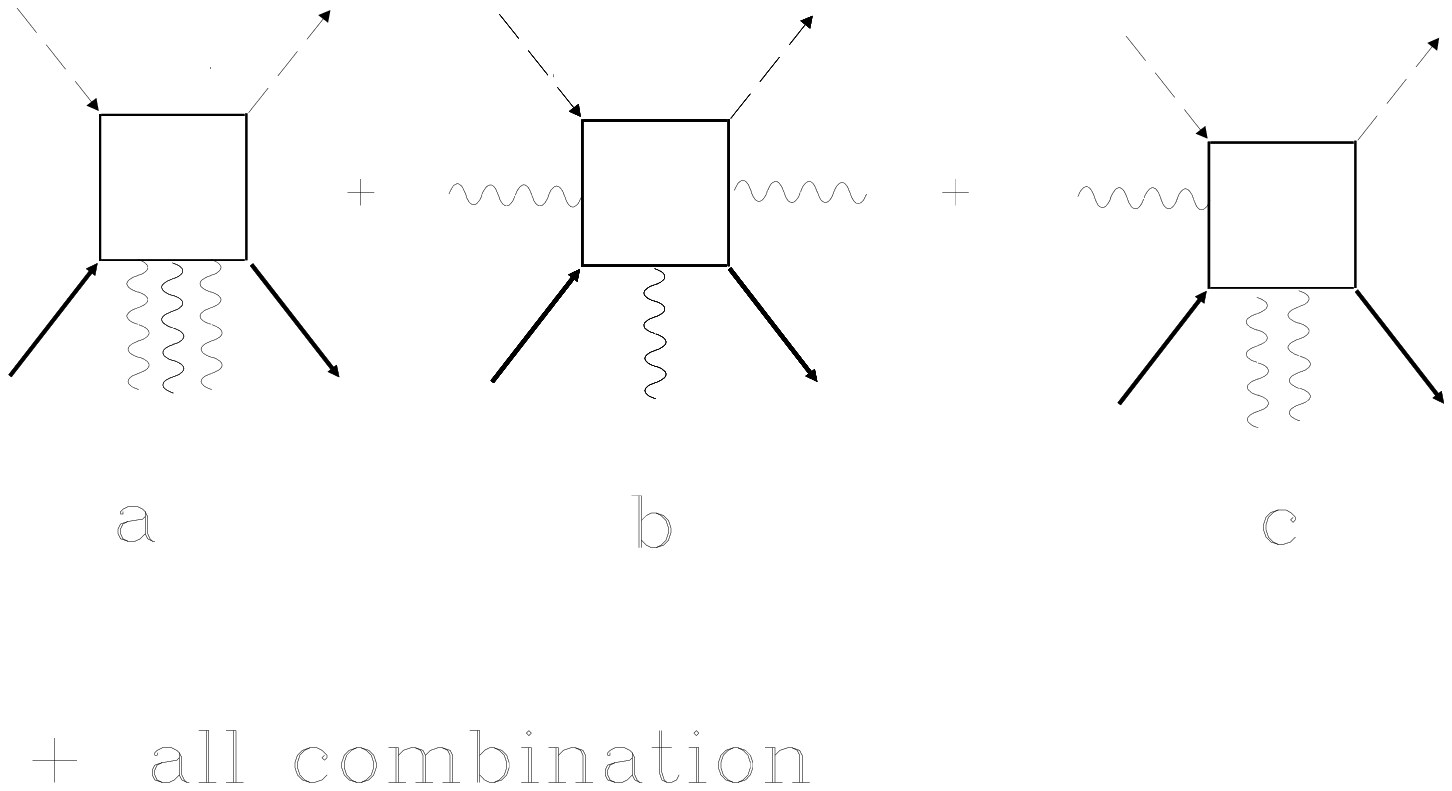} \caption{diagrams, corresponding
to $d=6$ condensate}
\end{figure}

\begin{figure}
\epsfxsize=5cm \epsfbox{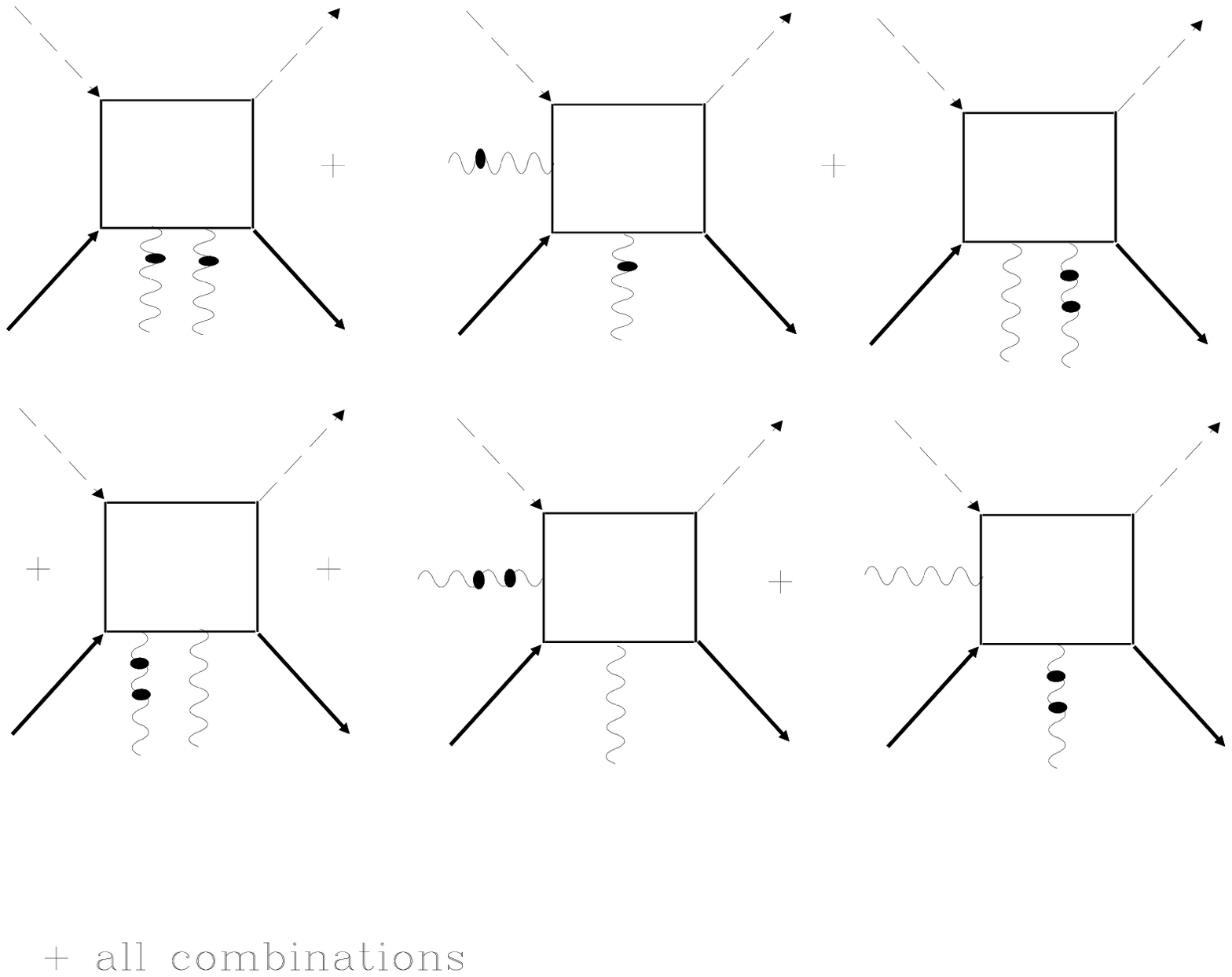} \caption{diagrams, corresponding
to $d=6$ condensate}
\end{figure}

\begin{figure}
\epsfxsize=5cm \epsfbox{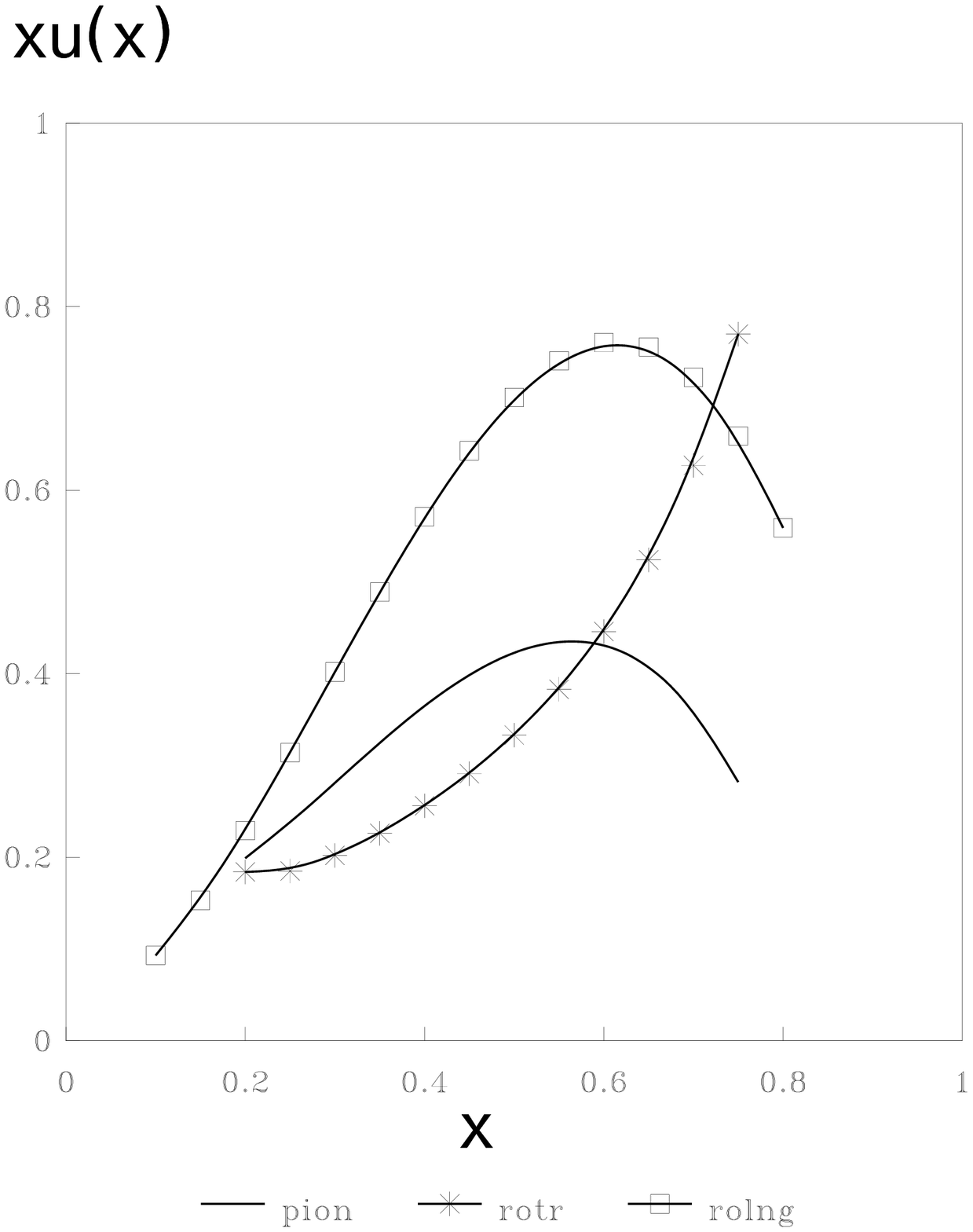} \caption{Quark distribution
function for pion, transversally and longitudinally polarized
$\rho$ - solid line, line with asters, and line with squares
correspondingly}
\end{figure}


\begin{thebibliography}{0}
\bibitem{1} V.M.Belyaev and B.L.Ioffe, Nucl. Phys. B 310, (1988) 548.
\bibitem{2} B.L.Ioffe and A.G.Oganesian, Eur. Phys. J. C 13 (2000) 485.
\bibitem{3} B.L.Ioffe and A.G.Oganesian, Phys. Rev. D 63 (2001) 096006.
\bibitem{4} B.L.Ioffe and A.Oganesian, Nucl. Phys. A 714 (2003) 145.
\bibitem{iz} B.L.Ioffe and K.N.Zyablyuk, Nucl. Phys. A 687 (2001) 437.
\bibitem{giz} B.V.Geshkenbein, B.L.Ioffe and K.N.Zyablyuk, Phys. Rev. D 64 (2001) 093009.
\bibitem{ii} B.L.Ioffe  Prog.Part.Nucl.Phys. 56 (2006) 232.
\bibitem{bb} V.Belyaev and B.Blok, Phys.Lett. B 167 (1986) 99,
Yad.Fiz. 43 (1986) 706.
\bibitem{is} B.L.Ioffe and A.V.Smilga,  Nucl.Phys. B 232 (1984), 109.
\bibitem{sam} A.Oganesian and A.Samsonov, JHEP 0109: 002, (2001).
\end{thebibliography}
\end{document}